# Quantum Photonic Circuits Integrated with Color Centers in Designer Nanodiamonds

Kinfung Ngan[1], Yuan Zhan[1], Constantin Dory[2], Jelena Vučković[2], and Shuo Sun[1*]

[1]JILA and Department of Physics, University of Colorado, Boulder, CO 80309, USA

[2]E. L. Ginzton Laboratory, Stanford University, Stanford, CA 94305, USA

[*]Email: shuosun@colorado.edu

**Abstract:** Diamond has emerged as a leading host material for solid-state quantum emitters, quantum memories, and quantum sensors. However, the challenges in fabricating photonic devices in diamond have limited its potential for use in quantum technologies. While various hybrid integration approaches have been developed for coupling diamond color centers with photonic devices defined in a heterogeneous material, these methods suffer from either large insertion loss at the material interface or evanescent light-matter coupling. Here, we present a new technique that enables deterministic assembly of diamond color centers in a silicon nitride photonic circuit. Using this technique, we observe Purcell enhancement of silicon vacancy centers coupled to a silicon nitride ring resonator. Our hybrid integration approach has the potential for achieving the maximum possible light-matter interaction strength while maintaining low insertion loss, and paves the way towards scalable manufacturing of large-scale quantum photonic circuits integrated with high-quality quantum emitters and spins.

Diamond is a unique host material for a diverse range of color centers with remarkable optical and spin coherence.[1,2] This is largely due to its wide bandgap, exceptional chemical stability, and the feasibility of synthesizing single-crystal diamond with extremely low impurities at parts-per-billion level.[3] Groundbreaking proof-of-concept experiments have been reported by employing diamond color centers as single-photon emitters or optically accessible spin qubits, including single-photon generation,[4] entanglement distribution and swapping,[5-7] quantum teleportation,[8] and memory-enhanced quantum communication.[9,10] To improve the performance and scalability of these applications requires the integration of diamond color centers with scalable photonic circuits.[11] However, despite significant advancements in diamond fabrication in the past decade,[12-15] creating large-scale photonic circuits in diamond remains a significant challenge. Additionally, the stochastic nature of color center creation makes it difficult to achieve scalable and deterministic integration of color centers with photonic circuits.

Hybrid integration, the process of integrating quantum emitters with photonic circuits in a heterogeneous material, provides a viable solution to both aforementioned challenges.[16,17] Hybrid integration allows us to take the best of both worlds, the high-quality quantum emitters and spins based on color centers in diamond, and the wafer-scale integrated photonic circuits in fabrication-friendly materials such as Silicon Nitride (SiN). In addition, one can use pre-selected quantum emitters for photonic integration, making hybrid integration a scalable way to assemble complex quantum photonic circuits. Two different approaches have been developed for hybrid integration of diamond color centers with heterogeneous photonic circuits. The first one relies on the pick-and-place technique,[18,19] where diamond photonic devices containing color centers are picked up from a diamond substrate and placed onto sockets defined on a heterogeneous photonic circuit. This approach allows for the maximum possible coupling between the color center and the diamond photonic device such as a waveguide or a cavity. However, it suffers from a relatively large insertion loss (typically a few dB)[19] at the interface between the diamond waveguide and the photonic circuit, due to a combination of placement errors and photon scattering. The second approach relies on evanescent coupling of diamond color centers with photonic devices defined completely in the heterogeneous material.[20-26] This can be achieved by fabrication of photonic devices in a high-refractive-index material deposited on single-crystal diamond,[20-23] or by placing nanodiamonds on top of a photonic device defined in a heterogeneous material.[23-26] While this approach avoids insertion loss at the material interface, it suffers from weak evanescent coupling between the quantum emitter and the photonic device.

In this Letter, we report a new hybrid integration approach that has the potential to achieve the maximum possible light-matter interaction strength while maintaining low insertion loss at the material interface. This approach is achieved by deterministic assembly of nanodiamonds *inside* Silicon Nitride (SiN), followed by fabrication of photonic devices in SiN aligned with the nanodiamonds. Using the new hybrid integration technique, we demonstrate deterministic integration of diamond Silicon Vacancy (SiV) centers with photonic waveguides and cavities defined in SiN. We also observe Purcell enhancement of SiV centers coupled to a SiN ring resonator. Our hybrid integration technique is not specific to SiV centers or SiN photonics. It is generally applicable to the integration of virtually any solid-state quantum emitters with photonic devices defined in thin-film materials. Our results thus represent an important step towards scalable integration of solid-state quantum emitters with large-scale integrated photonics.

Our hybrid integration approach started with top-down fabrication of nanodiamonds based on electron beam lithography (see Supplementary Section 1).[27] Compared with other methods such as High-Pressure-High-Temperature (HPHT) synthesis,[28] top-down fabrication of nanodiamonds enables precise control of the size and shape of each nanodiamond, which is crucial for scalable photonic integration. For this reason, we refer these nanodiamonds as the designer nanodiamonds. We started with an electronic grade single-crystal diamond from Element Six. Following standard diamond pretreatment based on tri-acid clean and oxygen plasma etching, the diamond sample was implanted uniformly with $^{28}$Si$^+$ ions at CuttingEdge Ions, and then annealed under vacuum. This process created a uniform distribution of SiV centers inside the diamond, with a density that can be controlled by the implantation dose. On this sample, we fabricated an array of nanopillars with a diameter of ~150 nm and a height of ~400 nm via electron beam lithography. We then performed partial undercut of the nanopillars by using quasi-isotropic etching of diamond.[13-15] This step created designer nanodiamonds in the shape of a small cylinder (~130 nm long) sitting on top of a thin neck (~10 nm thin). Figures 1a and 1b show the Scanning Electron Microscope (SEM) images of the fabricated designer nanodiamonds viewed from the top (Fig. 1a) and the side (Fig. 1b), respectively.

The partial undercut is essential, as it allows us to deterministically break each designer nanodiamond off from the diamond substrate and assemble them one by one on a heterogeneous substrate. In this work, we used a nanoprobe from an SEM integrated Focused Ion Beam (FIB) system for the transfer of designer nanodiamonds, which allowed us to position the designer nanodiamonds while imaging their positions using the SEM at the same time. As a proof of concept, we were able to assemble three designer nanodiamonds in a line on the thin-film SiN (see Fig. 1c for the SEM image).

To examine the properties of the SiV centers inside the designer nanodiamonds, we measured the photoluminescence excitation spectra of 13 SiV centers in 9 designer nanodiamonds transferred on a thin-film SiN. Figure 1d shows the histogram of the measured linewidth of the 13

SiV centers (see Supplementary Section 2 for the photoluminescence excitation spectrum of each emitter). We measured a linewidth of 290(10) MHz for the narrowest SiV center and a statistical average linewidth of 460(30) MHz for the 13 SiV centers we measured. The average linewidth is 5 times the lifetime-limited value (94 MHz), which could be due to both implantation-induced lattice damage and the proximity to surfaces. This number is comparable to SiV centers monolithically integrated in a diamond photonic crystal, where a statistical average linewidth of 410(160) MHz was reported.[29] Figure 1e shows the histogram of the range of spectral diffusion in 10 minutes for the same 13 SiV centers. The majority of the SiV centers had a range of spectral diffusion less than 500 MHz, which is also comparable to SiV centers monolithically integrated in a diamond photonic crystal.[29,30] The similarities of the optical properties between SiV centers in designer nanodiamonds and SiV centers monolithically integrated in diamond nanostructures were expected as the average distance from the SiV center to the surface of the designer nanodiamond is similar to SiV centers in a diamond photonic crystal.[29,30] The spectral stability of the SiV centers could potentially be further improved by Hydrogen plasma treatment of the diamond surface, a technique that was shown effective on SiV centers in HPHT-synthesized nanodiamonds.[31]

The capability in assembling designer nanodiamonds on thin-film SiN enables deterministic hybrid integration of color centers in designer nanodiamonds with SiN nanophotonic devices. We started the hybrid integration by depositing a thin layer of SiN on top of a $SiO_2$ buffer layer. Following deterministic placement of designer nanodiamonds at the desired locations defined by metal markers patterned on the SiN layer, we deposited a second layer of SiN. This step guarantees that the designer nanodiamonds are buried inside the SiN photonic layer, which enables maximum possible coupling between the SiV centers and the SiN photonic devices. We then fabricated nanophotonic devices in SiN centered around each designer nanodiamond. We refer the readers to Supplementary Section 1 for details of the hybrid integration process.

We first demonstrate photonic coupling between a SiV center and a SiN nanobeam waveguide in a photonic circuit. Figure 2a shows the SEM image of a simple SiN photonic circuit integrated with a single designer nanodiamond (as indicated by the small bump shown in the inset SEM image in Fig. 2a). The SiN photonic circuit consisted of a nanobeam waveguide connected to a 50/50 power splitter. Each output port of the power splitter was coupled to free space via a grating coupler. It is noteworthy that despite the simplicity of this photonic circuit, fabrication of the same device in diamond for monolithic integration is extremely difficult due to the complex dimensions of the device structure. In contrast, fabrication of this device in SiN is straightforward, and we can follow the hybrid integration process described earlier to incorporate a single designer nanodiamond containing SiV centers into the center of the SiN waveguide, as verified by the scanning photoluminescence microscopy image (the top inset of Fig. 2a). To confirm waveguide-emitter coupling, we directly excited the designer nanodiamond from the top via free space with a green laser, and collected the photoluminescence of the SiV center from both the top via free space and the two output ports of the 50/50 power splitter. The photoluminescence spectra of both output ports (Figs. 2b and 2c) exhibited clear spectral signature of the four zero-phonon lines of the SiV center[32] and were nearly identical with the photoluminescence spectrum observed from the top (Fig. 2d), demonstrating photonic coupling between the SiV center and the SiN waveguide.

To examine whether the hybrid integration process degraded the properties of the SiV center, we measured the photoluminescence excitation spectra of a specific optical transition of the SiV center both before (Fig. 2e) and after (Fig. 2f) the hybrid integration. This specific transition connects the lower excited state and the lower ground state of the SiV center, and is conventionally referred to as the C transition. We observed a reduced linewidth from 490(30) MHz to 360(4) MHz. We also measured the range of spectral diffusion by monitoring the photoluminescence excitation spectra of the SiV center over a 10-minute interval, both before (Fig. 2g) and after (Fig. 2h) the hybrid integration process, and observed a qualitatively smaller range of spectral diffusion after the hybrid integration. We attribute the improvement in the optical properties of the SiV center to a more stabilized charge environment at the surface of the designer nanodiamond after the hybrid integration. In particular, we observed significant improvement in the emitter linewidth, emitter blinking, and the range of spectral diffusion for multiple SiV centers after the hybrid integrated sample was at rest for a long time (see Supplementary Section 3 for details). This observation suggests that the charge environment of the sample near the designer nanodiamond is crucial to maintain the optical coherence of the embedded SiV center.

Building on the capability of deterministic hybrid integration, we now demonstrate Purcell enhancement by hybrid integration of a SiV center with a SiN ring resonator. Figure 3a shows the SEM image of the SiN ring resonator integrated with a designer nanodiamond. The arrow in Fig. 3a indicates the position of the designer nanodiamond. To observe the Purcell effect, we measured the photoluminescence spectra of the SiV center while tuning the cavity resonances via gas condensation (see Supplementary Section 5 for details of the measurement). Figure 3b shows the measured photoluminescence spectra as a function of time, and the white dashed lines in Fig. 3b indicate the resonances of each cavity mode. We observed a 2-time enhancement of emission intensity from the C transition of the SiV center when a cavity mode is resonant with this transition, suggesting cavity induced Purcell enhancement.

To further verify the Purcell effect, we measured the same SiV center before and after its integration with the cavity, and compared the measured lifetime directly. Figures 3c-3e show the time-resolved photoluminescence of the SiV center before (Fig. 3c) and after (Figs. 3d and 3e) its integration with the SiN ring resonator. In Fig. 3d, the cavity modes are off resonant with any of the four zero-phonon lines, and in Fig. 3e, one of the cavity modes is resonant with the C transition of the SiV center. The lifetime of the SiV center remained unchanged before and after integration when the cavity modes were off resonant, but was reduced by ~6.5% (5 times of the measurement uncertainty) when a cavity mode was in resonance with the SiV center. We estimated a lower bound of the Purcell factor of 2 by considering a higher bound of the SiV quantum efficiency of 30%, the zero-phonon-line fraction of 80%, and a fraction of 30% of zero-phonon-line emission into the C transition that couples with the cavity (see Supplementary Section 6 for details of the calculation). The measured Purcell factor is 1/4 of the theoretically attainable value (see Supplementary Section 7). We attribute the mismatch to the position and polarization misalignment between the emitter and the cavity mode.

A legitimate concern of our hybrid integration approach is that the integration of the designer nanodiamond may degrade the Q factor of the SiN cavity. The degradation in cavity Q factor can be possibly caused by either the imperfect fabrication in hybrid integration (e.g. the generation of air gaps between the designer nanodiamond and SiN) or the refractive index mismatch between the diamond and the SiN. Here we use $Q_{hy}$ to characterize the Q factor limited by cavity loss induced by the hybrid integration, which satisfies $\frac{1}{Q_{hy}} = \frac{1}{Q_{loaded}} - \frac{1}{Q_{unloaded}}$, where $Q_{loaded}$ and $Q_{unloaded}$ are the Q factors of the cavity with and without the integrated designer nanodiamond, respectively. To experimentally find $Q_{hy}$, we fabricated two identical cavities on the same wafer by going through the same fabrication run. The only difference between the two cavities was that one cavity was integrated with a designer nanodiamond, whereas the other was not. Figures 4a and 4b show the measured transmission spectra of the cavity with (Fig. 4a) and without (Fig. 4b) the integrated designer nanodiamond. From the cavity transmission spectra, we measured $Q_{loaded} = (1.01 \pm 0.06) \times 10^4$ and $Q_{unloaded} = (1.17 \pm 0.09) \times 10^4$. Based on these measurements, we calculated $Q_{hy}$ to be $Q_{hy} = (7 \pm 5) \times 10^4$.

To further investigate the origin of loss due to hybrid integration, we numerically simulated the Q factor of the loaded cavity ($Q_{loaded}$) as we varied the Q of the unloaded cavity ($Q_{unloaded}$). Here, we varied $Q_{unloaded}$ by changing the gap distance between the cavity and a bus waveguide. In our simulation, we assumed the designer nanodiamond to be a perfect cylinder, with a diameter of 150 nm and a height of 130 nm, which resembles the shape of the designer nanodiamond used in our experiments. In addition, we assumed no air gaps between the designer nanodiamond and the SiN material. Figure 4c shows the calculated value of $Q_{loaded}$ as a function of the unloaded cavity Q factor $Q_{unloaded}$. By fitting the simulated data to an analytical function of $\frac{1}{Q_{loaded}} = \frac{1}{Q_{hy}} + \frac{1}{Q_{unloaded}}$, we obtained that $Q_{hy} = (4.7 \pm 0.3) \times 10^4$. This value matches with our experimentally measured value, suggesting that the main loss induced by hybrid integration is due to the refractive index mismatch between the designer nanodiamond and the SiN material. We can potentially reduce the hybrid-integration-induced loss by using smaller designer nanodiamonds or by using a heterogeneous material that has a refractive index closer to diamond (e.g. $TiO_2$).[33]

In summary, we reported a new approach for scalable assembly and hybrid integration of diamond color centers with SiN nanophotonic circuits. Our approach has two main potential advantages. First, it is scalable since we can pre-characterize each designer nanodiamond and integrate only the desired ones with the photonic device. Second, our approach allows for the maximum possible emitter-light coupling strength and low insertion loss at the material interface. With our current cavity device, when the SiV center is in its excited state, the probability of emitting a photon into a single SiN cavity mode is 6.5%. By employing improved nanofabrication techniques and utilizing smaller designer nanodiamonds, we can improve the photon extraction probability to beyond 97% (see Supplementary Section 8). It should be noted that nanodiamonds of smaller sizes may adversely affect the coherence properties of the color centers. Therefore, one may need to develop additional surface treatment techniques[31, 34] for maintaining the coherence of the color center. The use of cavities with smaller mode volumes may further boost this efficiency if the cavity is designed properly such that the integrated designer nanodiamond does not post a smaller ceiling on the cavity Q factor. In Supplementary Section 8, we presented a specific design of a fishbone photonic crystal cavity that serves for this purpose. The devices reported in this work contain only one designer nanodiamond each. However, the hybrid integration approach we developed here can be easily extended for the deterministic integration of multiple designer nanodiamonds with the same photonic device. This platform is uniquely suited for studying many-body quantum interactions in an engineerable photonic bath.[35-36] Ultimately, combined with recent progresses in foundry-based large-scale integrated photonics in SiN,[37-39] our hybrid integration approach paves the way towards scalable manufacturing of quantum photonic circuits integrated with high-quality quantum emitters and spins.

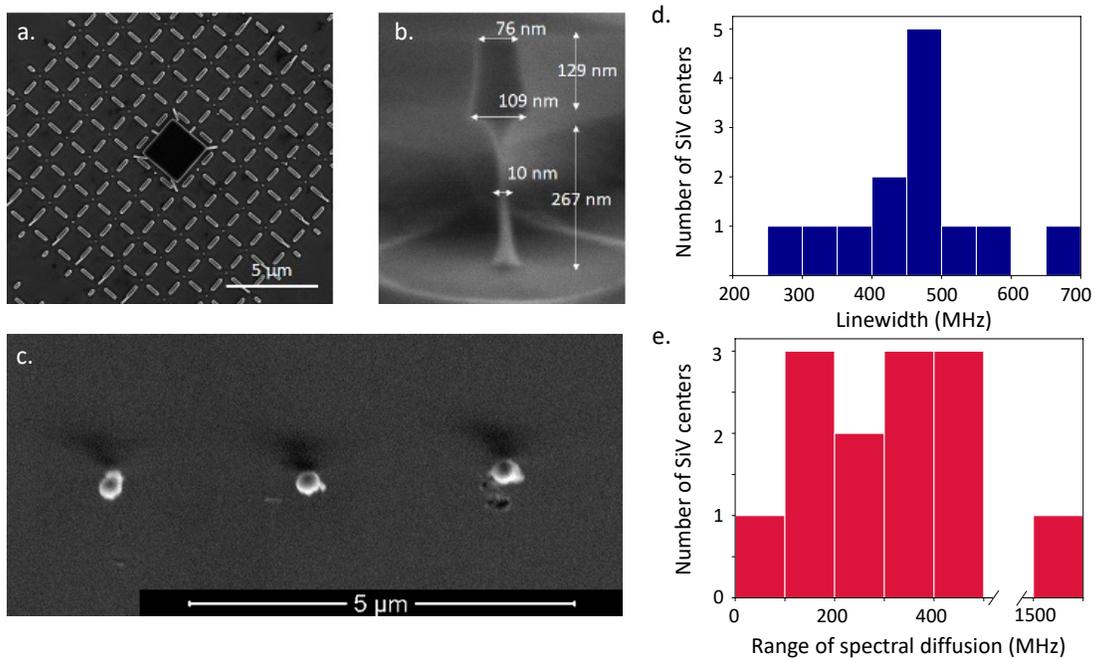

**Figure 1.** Fabrication, assembly, and characterization of designer nanodiamonds. (a) Scanning Electron Microscope (SEM) image of an array of designer nanodiamonds fabricated on a single-crystal diamond substrate. (b) SEM image of the side view of a single designer nanodiamond. (c) SEM image of three designer nanodiamonds deterministically assembled on a thin-film SiN arranged in a line. (d) Histogram of the linewidth of 13 SiV centers in 9 designer nanodiamonds transferred from the diamond substrate to the thin-film SiN. (e) Histogram of the range of spectral diffusion in 10 minutes of the same 13 SiV centers in 9 transferred designer nanodiamonds.

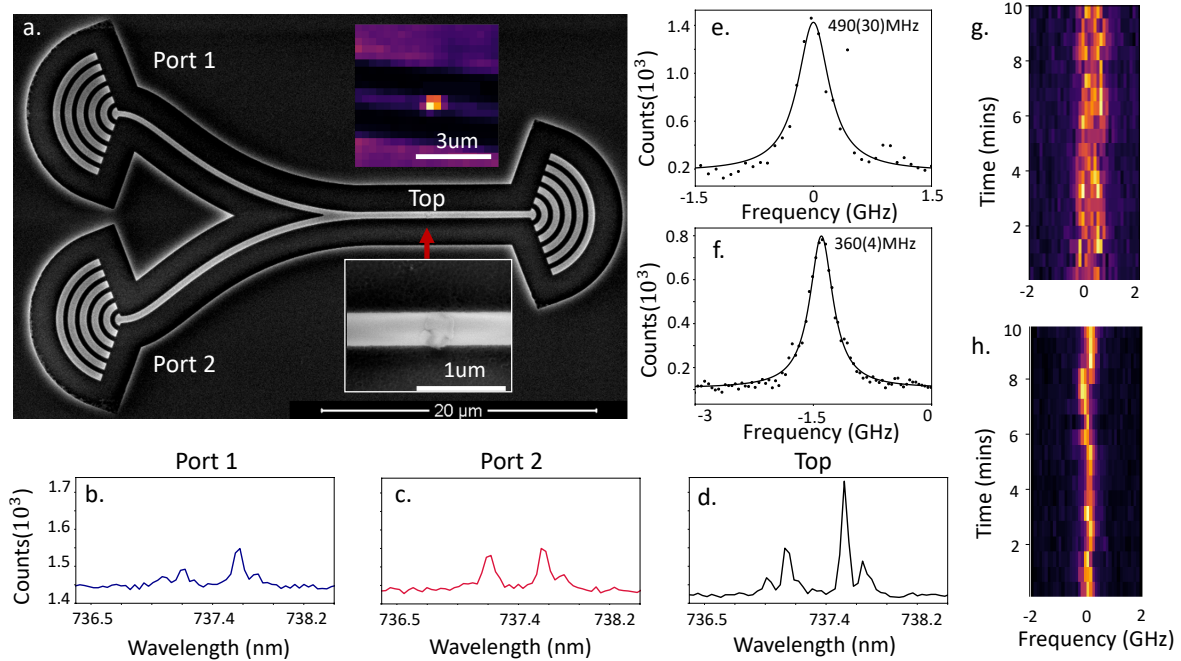

**Figure 2.** Deterministic photonic coupling between a SiV center in a designer nanodiamond and SiN photonic circuits. (a) SEM image of a SiN photonic circuit integrated with a single designer nanodiamond. The photonic circuit consists of a single nanobeam waveguide coupled to a 50/50 power splitter. The designer nanodiamond is integrated inside the nanobeam waveguide. The bottom inset is an SEM image of a zoom-in image of the area centered around the designer nanodiamond. The top insert shows the scanning confocal photoluminescence image of the area centered around the designer nanodiamond. (b-d) Photoluminescence spectra of the SiV center collected from output port 1 (b), output port 2 (c), and the top of the integrated designer nanodiamond via free space (d). (e, f) Photoluminescence excitation spectra of the same optical transition from the same SiV center before (e) and after (f) the integration of the designer nanodiamond with the SiN photonics. The black dots show the measured data, and the black solid lines show the Lorentzian fits to the measured data. (g, h) Photoluminescence excitation spectra of the same optical transition shown in (e) and (f) recorded for 10 minutes before (g) and after (h) the integration of the designer nanodiamond with the SiN photonics. The designer nanodiamond was excited via free space for all spectra.

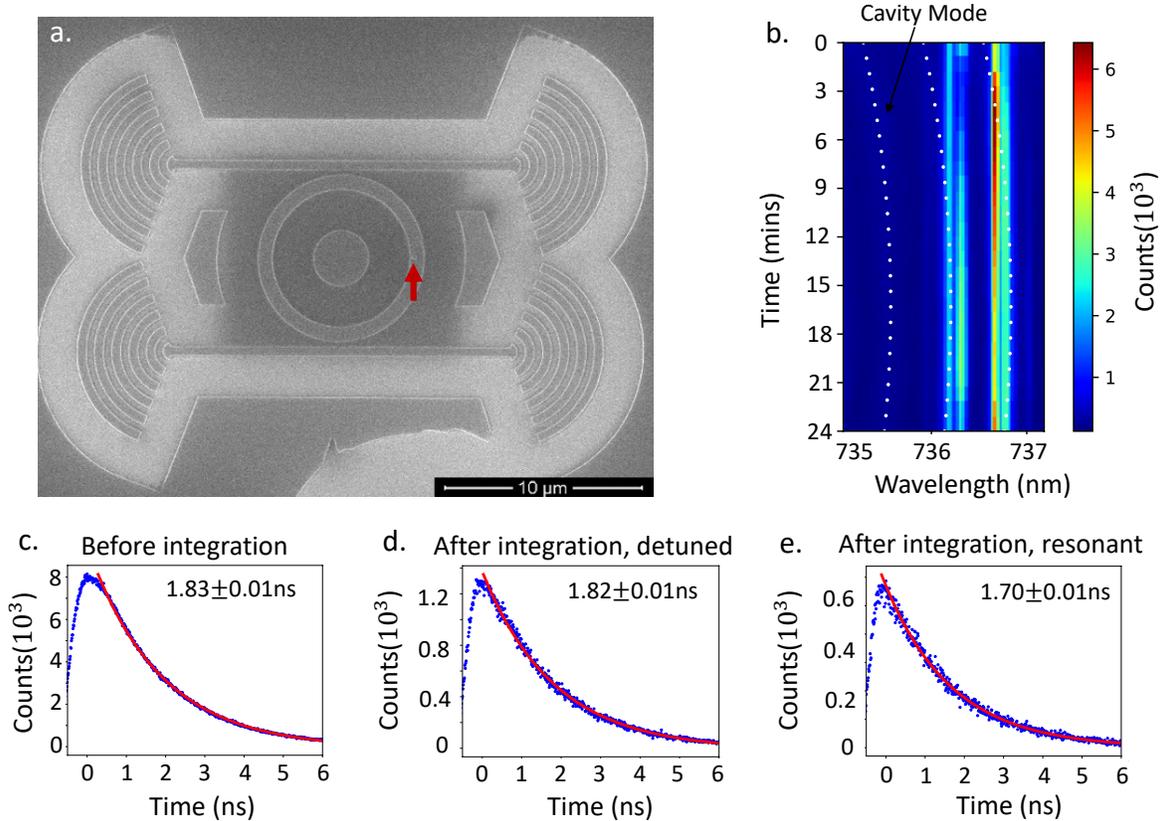

**Figure 3.** Purcell enhancement of a SiV center induced by a SiN ring resonator. (a) SEM image of the SiN ring resonator integrated with a single designer nanodiamond. The red arrow indicates the position of the integrated designer nanodiamond. (b) Photoluminescence spectra of the SiV center as a function of time. The white dashed lines indicate the resonance of each cavity mode as a function of time (see Supplementary Section 5 for how we retrieved the cavity resonances). (c-e) Time-resolved photoluminescence of the SiV center before the integration (c), after the integration but with all cavity modes far detuned from any of the four zero-phonon lines (d), and after the integration and with one cavity mode resonant with the transition connecting the lower excited state and the lower ground state (e). The blue dots are the measured data, and the red solid lines are the numerical fits to an exponential decay.

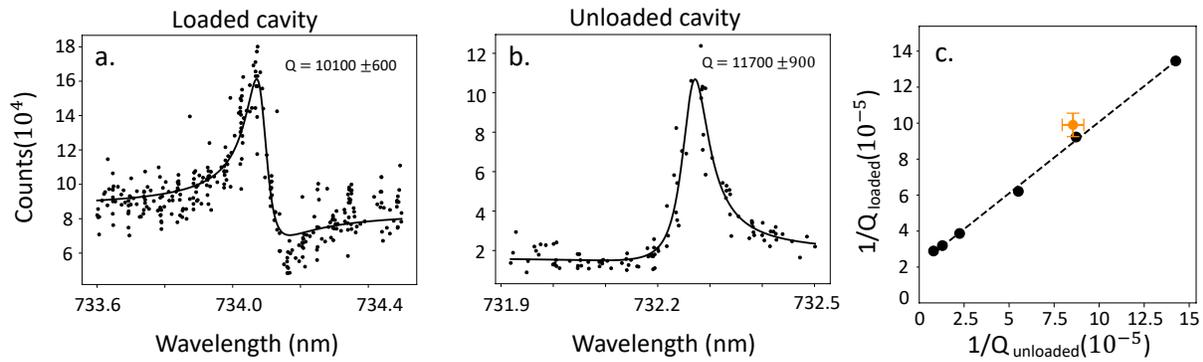

**Figure 4.** Cavity loss introduced by hybrid integration. (a, b) Transmission spectra of the loaded (a) and unloaded (b) cavity. The black dots are the measured data, and the black solid lines are Lorentzian fits of the measured data, from which we derived the Q factor. (c) Calculated values of $Q_{loaded}$ as a function of $Q_{unloaded}$. Here we varied the value of $Q_{unloaded}$ by changing the coupling between the cavity and a bus waveguide. The black dots show the calculated values of $Q_{loaded}$ and $Q_{unloaded}$ for each waveguide-cavity coupling strength, and the black dashed line shows a linear fit. The orange dot indicates the experimentally measured value.

**Corresponding Author**

*Shuo Sun: shuosun@colorado.edu

**Supporting Information**

Details of nanofabrication, experiments, simulation, and additional supporting data and analysis.

**Funding Sources**

Funding for this work is provided by the National Science Foundation (NSF) (2032567, 2150633, 2317149, and 2326628), the Air Force Office of Scientific Research (AFOSR) (FA2386-21-1-4084), the W. M. Keck Foundation, and the Research & Innovation Office (RIO) Core Facility Assistance Grant Program at the University of Colorado Boulder. Device fabrication was performed, in part, at the Center for Integrated Nanotechnologies, an Office of Science User Facility operated for the U.S. Department of Energy (DOE) Office of Science by Los Alamos National Laboratory (Contract 89233218CNA000001) and Sandia National Laboratories (Contract DE-NA-0003525). The pick-and-place of designer nanodiamonds was performed with the facility at the Colorado Shared Instrumentation in Nanofabrication and Characterization (COSINC): the COSINC-CHR (Characterization) and/or CONSINC-FAB (Fabrication), College of Engineering & Applied Science, and the University of Colorado Boulder. The authors would like to acknowledge the support of Tomoko Borsa and the facility that have made this work possible. The designer nanodiamonds sample was fabricated at the Stanford Nanofabrication Facility (SNF) and the Stanford Nano Shared Facilities (SNSF), supported by the National Science Foundation under award ECCS-1542152. Shuo Sun acknowledges support from the Sloan Research Fellowship.

## Acknowledgements

The authors acknowledge Andrew Mounce for fruitful discussions and Yijun Xie for advice on nanofabrication. The authors acknowledge funding from the National Science Foundation (NSF) (2032567, 2150633, 2317149, and 2326628), the Air Force Office of Scientific Research (AFOSR) (FA2386-21-1-4084), the W. M. Keck Foundation, and the Research & Innovation Office (RIO) Core Facility Assistance Grant Program at the University of Colorado Boulder. Device fabrication was performed, in part, at the Center for Integrated Nanotechnologies, an Office of Science User Facility operated for the U.S. Department of Energy (DOE) Office of Science by Los Alamos National Laboratory (Contract 89233218CNA000001) and Sandia National Laboratories (Contract DE-NA-0003525). The pick-and-place of designer nanodiamonds was performed with the facility at the Colorado Shared Instrumentation in Nanofabrication and Characterization (COSINC): the COSINC-CHR (Characterization) and/or CONSINC-FAB (Fabrication), College of Engineering & Applied Science, and the University of Colorado Boulder. The authors would like to acknowledge the support from Tomoko Borsa and the facility that has made this work possible. The designer nanodiamonds sample was fabricated at the Stanford Nanofabrication Facility (SNF) and the Stanford Nano Shared Facilities (SNSF), supported by the National Science

Foundation under award ECCS-1542152. Shuo Sun acknowledges support from the Sloan Research Fellowship.

**Competing interests:** The authors declare no competing interests.

**TOC Graphic**

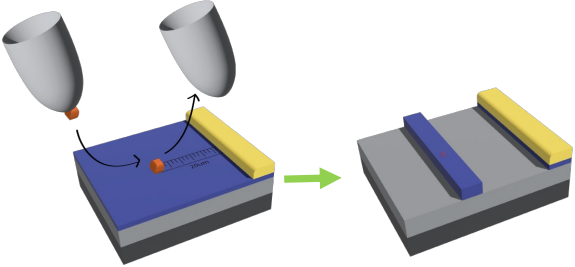

# Supplementary Information: Quantum Photonic Circuits Integrated with Color Centers in Designer Nanodiamonds

Kinfung Ngan[1], Yuan Zhan[1], Constantin Dory[2], Jelena Vučković[2], and Shuo Sun[1*]

[1]JILA and Department of Physics, University of Colorado, Boulder, CO 80309, USA

[2]E. L. Ginzton Laboratory, Stanford University, Stanford, CA 94305, USA

*Email: shuosun@colorado.edu

## 1. Process for diamond fabrication and hybrid integration

Figure S1 summarizes the steps we took for the fabrication of the designer nanodiamonds. We started our fabrication with a single crystal diamond containing randomly positioned SiV centers. We first defined a SiN hard mask on the diamond (Fig. S1a). This step was achieved by the deposition of a 200-nm-thick SiN via Low Pressure Chemical Vapor Deposition (LPCVD), followed by spin coating of the photoresist Hydrogen SilsesQuioxane (HSQ), followed by standard electron beam lithography to define the etching pattern on the HSQ and Reactive Ion Etching (RIE) ($SF_6$, $CH_4$, and $N_2$) to transfer the pattern from HSQ to SiN. We then transferred the pattern from SiN to the diamond via a directional RIE ($O_2$) (Fig. S1b). After this step, we used Atomic Layer Deposition (ALD) to coat ~20 nm Alumina isotopically on the sample, and then removed all of them except for those on the side wall via a directional RIE ($Cl_2$, $BCl_2$, and $N_2$) (Fig. S1c). After another directional RIE ($O_2$) of diamond to expose the pillar sidewalls below the designer nanodiamond (Fig. S1d), the sample was ready for undercut. As illustrated in Fig. S1e, the diamond undercut was achieved via quasi-isotropic etching of diamond. After igniting the oxygen plasma in RIE using forward bias, we turned off the forward bias and operated at 3000W inductively coupled plasma and elevated temperatures. The etch was faster along the <110> (in-plane direction) compared to the <001> (out-of-plane direction) crystal axis, resulting in an undercut of diamond pillars. We finally soaked the sample in hydrofluoric acid to remove the SiN hard mask and the Alumina side walls (Fig. S1f).

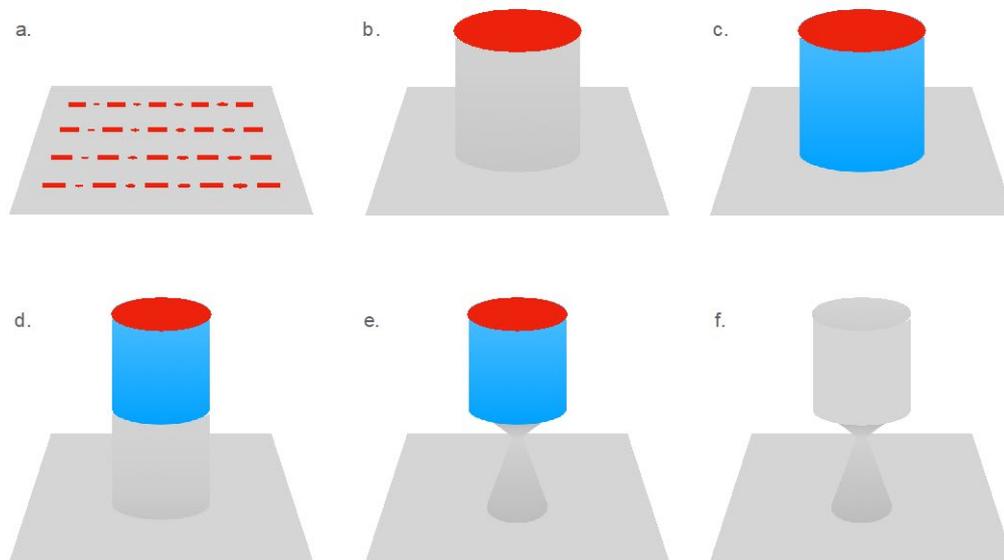

**Figure S1.** Schematic steps for fabrication of designer nanodiamonds.

Figure S2 summarizes the steps we took for hybrid integration. We first deposited a 1-μm-thick $SiO_2$ layer on a dummy Si substrate via LPCVD (Fig. S2a). We then deposited another 70 nm of SiN on top via Plasma Enhanced Chemical Vapor Deposition (PECVD) (Ar, $SiH_4$, $N_2$) (Fig. S2b). On this sample, we deposited a metal layer consisting of 10 nm Titanium and 100 nm Gold, and fabricated metal alignment markers via standard metal liftoff (Fig. S2c). We then transferred a designer nanodiamond to a target location defined by the alignment marker using a nanomanipulator (Ted Pella 460-102) under a dual FIB-SEM system (Fig. S2d). Following the transfer of the designer nanodiamond, we deposited a second layer of SiN of 230 nm via PECVD to cover the designer nanodiamond (Fig. S2e). This thickness was chosen to ensure that the expected position of the color center is at the center of the whole SiN layer deposited at both times. We then used standard electron beam lithography to fabricate photonic devices in the whole SiN layer, with the position of each photonic device determined by the metal alignment markers (Figs. S2f-S2h).

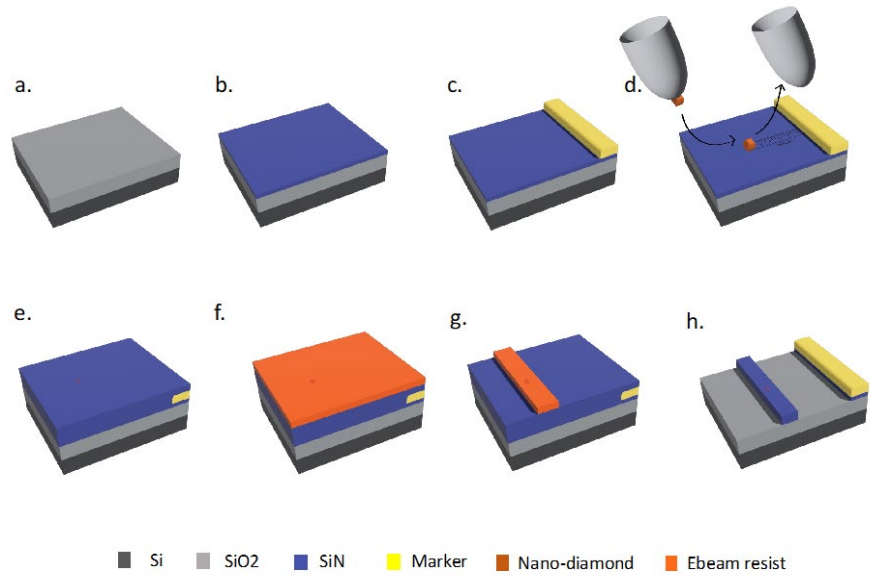

**Figure S2.** Schematic steps for hybrid integration between a color center in a designer nanodiamond and SiN photonic devices.

## 2. Characterization of SiV centers in designer nanodiamonds

In Fig. 1d of the main text, we presented the histogram of the linewidth of 13 different SiV centers in 9 designer nanodiamonds transferred onto a SiN substrate. The linewidth of each SiV center was measured via the photoluminescence excitation spectroscopy. Figure S3 shows the photoluminescence excitation spectra of the 13 SiV centers we measured, along with the extracted linewidth for each SiV center. These data were used to construct the histogram of the SiV linewidth shown in Fig. 1d of the main text.

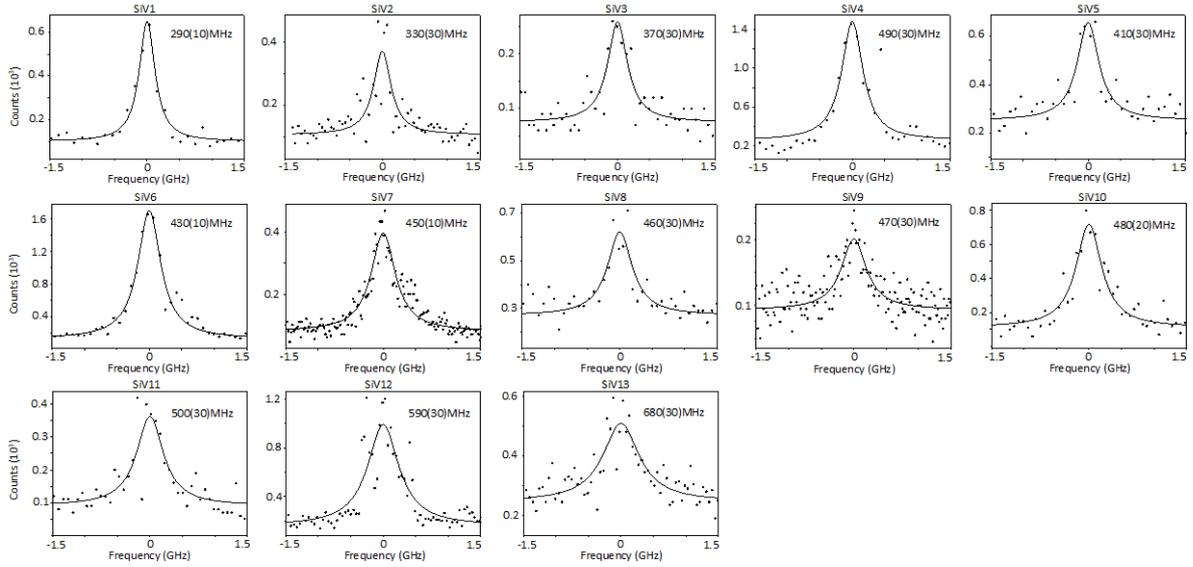

**Figure S3.** Photoluminescence spectra of 13 different SiV centers in 9 designer nanodiamonds transferred onto a SiN substrate.

## 3. Comparison of optical properties of SiV centers before and after hybrid integration

Figures S4a-S4c show the photoluminescence excitation spectra of the same SiV center presented in Fig. 2 of the main text but measured at three different times: before the hybrid integration (Fig. S4a, identical with Figs. 2e and 2g of the main text), right after the hybrid integration (Fig. S4b, not shown in the main text), and 6 months after the hybrid integration (Fig. S4c, identical with Figs. 2f and 2h of the main text). We observed similar linewidth of ~450 MHz both before and right after the hybrid integration. The range of spectral diffusion in 10 minutes was similar as well. However, the spectrum measured in Fig. S4b has a notably lower signal intensity and a higher background intensity than Fig. S4a. This is because the emitter was blinking right after the integration, and as a result a weak continuous-wave green laser was used to mitigate the emitter blinking. The photoluminescence induced by the weak green laser contributed to the higher background, whereas the remaining blinking of the emitter in the presence of the green laser contributed to the lower signal. Interestingly, after the sample was at rest for 6 months, we no longer observed the emitter blinking even in the absence of the green laser, and we were thus able to recover the signal-to-noise ratio of the spectrum. In addition, we observed a narrower emitter linewidth of 360 MHz and a qualitatively smaller range of spectral diffusion in 10 minutes. This result suggested that we were able to improve the optical properties of the emitter by having the sample at rest for a long time.

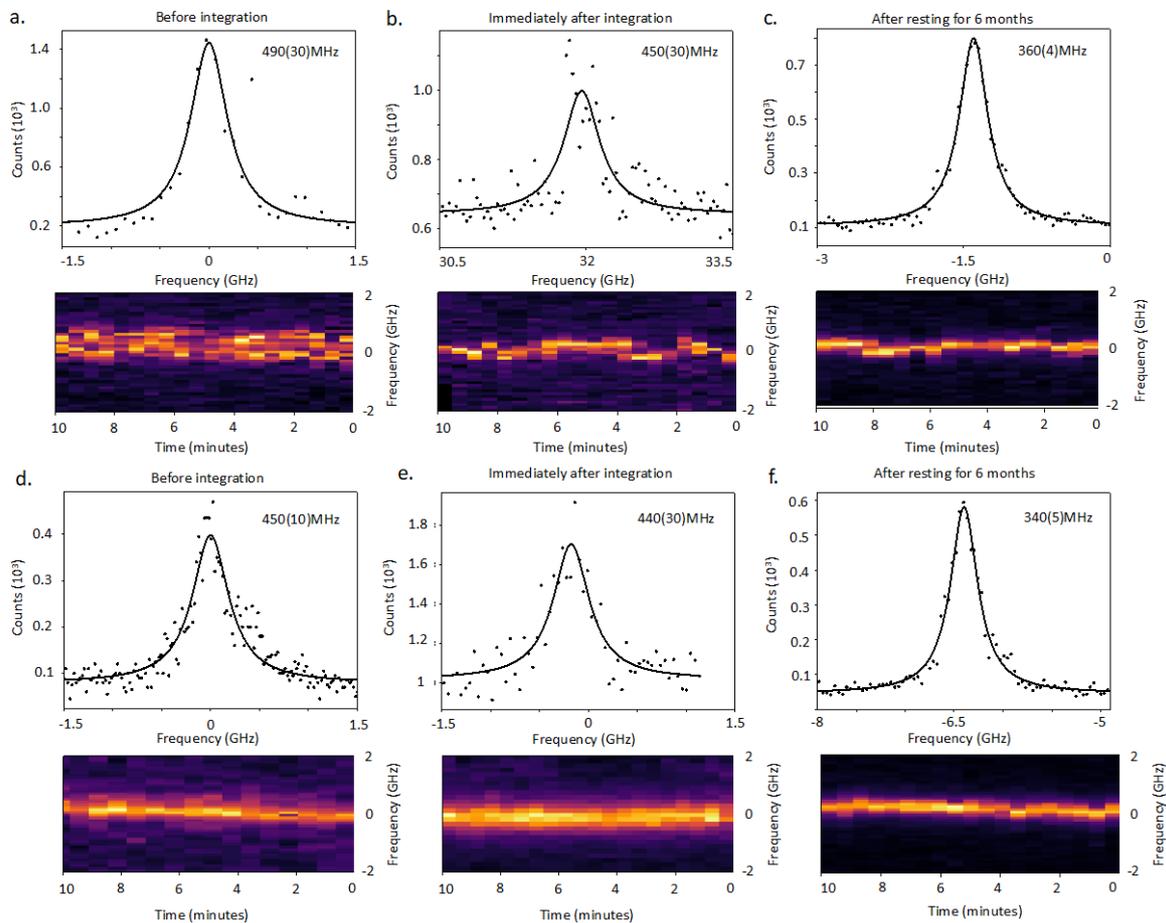

**Figure S4.** (a-c) Photoluminescence excitation spectra of the same SiV center shown in Fig. 2 of the main text, measured before the hybrid integration (a), right after the hybrid integration (b), and after the hybrid integrated sample was at rest for 6 months (c). (d-f) Photoluminescence excitation spectra of another SiV center in a different designer nanodiamond integrated with a different SiN device but on the same chip, measured before the hybrid integration (d), right after the hybrid integration (e), and after the hybrid integrated sample was at rest for 6 months (f). In all panels, the black dots show the measured data, the black solid lines show the Lorentzian fits to the measured data, and the heat maps show the spectra measured in multiple traces for 10 minutes.

To examine whether this effect applied to other emitters, we measured another SiV center in a different designer nanodiamond integrated with a different SiN device but on the same chip. Figures S4d-S4f show the measured photoluminescence excitation spectra of this SiV center measured before the hybrid integration (Fig. S4d), right after the hybrid integration (Fig. S4e), and after the sample was at rest for 6 months (Fig. S4f). Similar to the emitter shown in Figs. S4a-S4c, we observed emitter blinking right after the integration, which resulted in a lower signal-to-noise ratio in the photoluminescence spectrum (Fig. S4e). However, the emitter blinking was no longer present after the sample was at rest for 6 months. In addition, the linewidth of the emitter long time after hybrid integration also became narrower than both before integration and right after integration, consistent with what we observed with the SiV center shown in Figs. S4a-S4c.

We have also measured two more SiV centers in two different designer nanodiamonds integrated with different SiN devices on the same chip (Fig. S5). In both cases, the signal-to-noise ratio of the spectra were improved significantly after the sample was at rest for 6 months. In addition, the linewidth of the emitter became narrower as well. These results further confirmed that the optical properties of the SiV center in the hybrid device did improve after the sample was at rest for a long time. We note that in all of our measurements, we kept the same power for the excitation laser, which is well below the saturation threshold of the emitter.

We attribute the improvement of the SiV optical properties to a more stabilized charge environment after the sample was at rest for 6 months. Since we used the SEM to image the designer nanodiamond extensively right after the hybrid integration, the electrons in the SEM may accumulate at the surface of the designer nanodiamond, which contribute to the emitter blinking right after the hybrid integration. It is likely that these electrons were dissipated after the sample was at rest for a long time, leading to the significant mitigation of blinking and the narrower linewidth. It is not clear to us why the linewidth became even narrower than the same emitter before the hybrid integration. We conjectured that this may be due to the different surface property of the designer nanodiamond before and after hybrid integration. An interesting future direction is to perform more systematic study of how the surface property of the designer nanodiamonds affect the optical and spin coherence properties of the SiV centers inside, though it falls out of the scope of our current work.

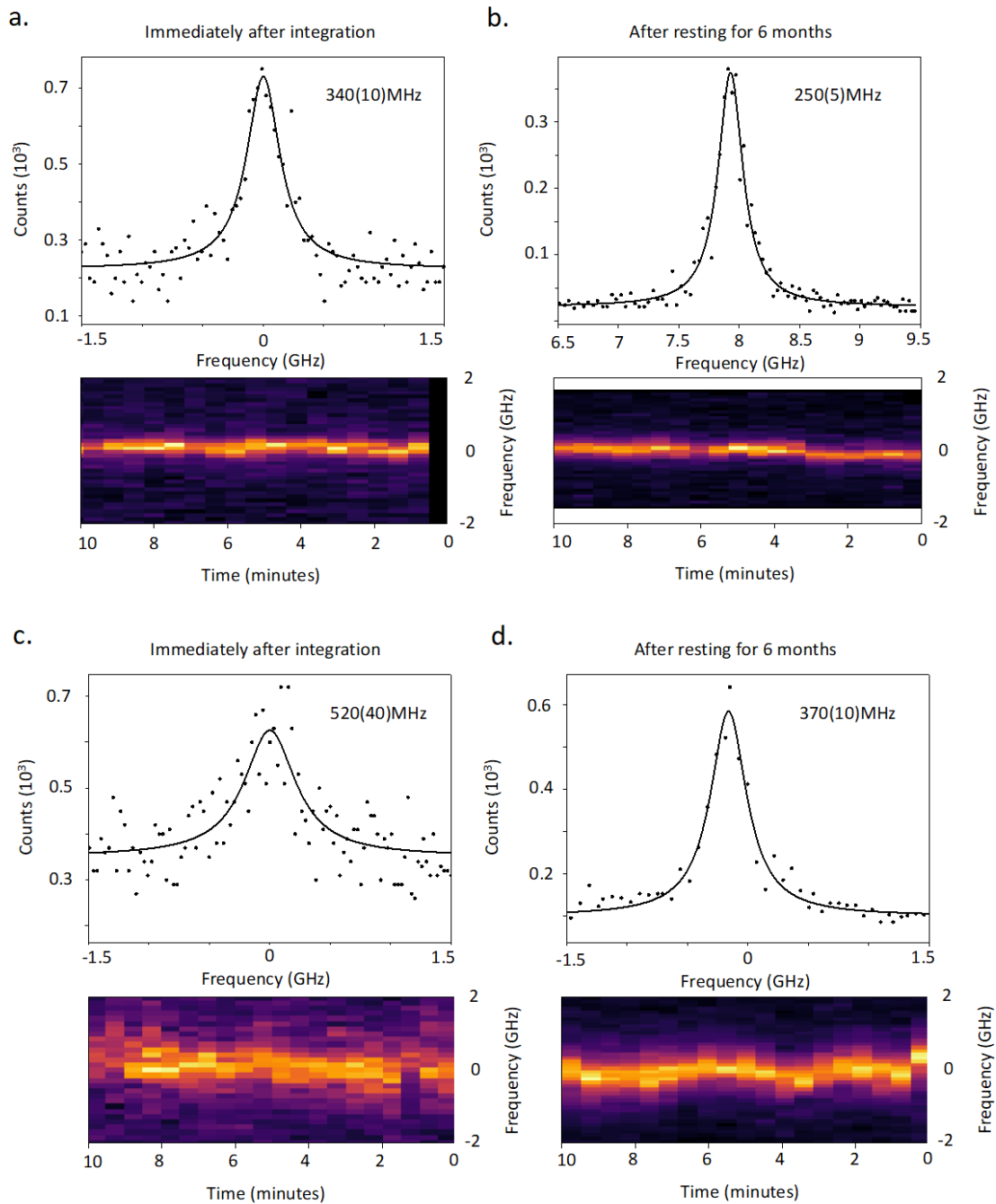

**Figure S5.** Photoluminescence excitation spectra of two more SiV centers inside two different designer nanodiamonds integrated with different SiN devices on the same chip, measured right after the hybrid integration (a, c) and after the sample was at rest for 6 months (b, d). In all panels, the black dots show the measured data, the black solid lines show the Lorentzian fits to the measured data, and the heat maps show the spectra measured in multiple traces for 10 minutes.

## 4. Calculation of collection efficiency of the photoluminescence of the waveguide-coupled SiV center

Figure S6 shows the calculated emission pattern of a dipole inside a designer nanodiamond coupled to a SiN waveguide. Based on the simulation, we calculated that 13% of photon emission was into the free space and can be collected by an objective lens with a numerical aperture of 0.9, and 20% of the emission was coupled to the waveguide.

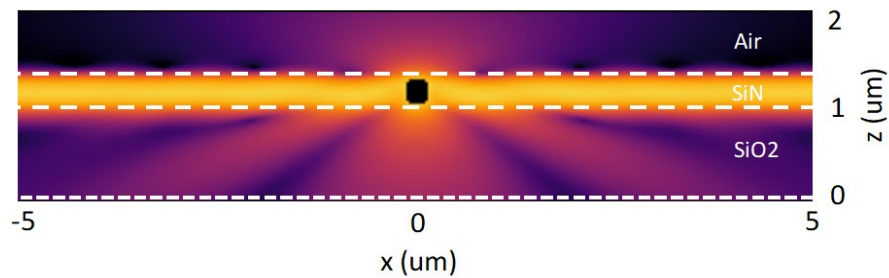

**Figure S6.** Emission pattern of a dipole inside a designer nanodiamond coupled to a SiN waveguide calculated by FDTD. The blacked center indicates the position of the dipole.

## 5. Measurements for cavity-induced Purcell enhancement

For Fig. 3b in the main text, to tune the cavity resonances, we employed gas condensation inside the cryostat, which happened by itself due to a minor vacuum leak in our cryostat. To initiate gas condensation, we started our experiment by warming up the whole cryostat to 40 K. At this temperature, the gas condensation on the sample was minimal. We then cooled down the chamber from 40 K to 4 K, which introduced gas condensation on the sample. We started recording the photoluminescence spectra of the SiV center when the sample reached about 4 K. At this point, even though the temperature of the sample was stabilized, there was still gas condensation going on. As gas condensed on the sample, it introduced redshift of the cavity resonances. At some point, when there was enough gas condensed on the sample, the cavity resonances no longer shifted, most likely due to the equilibrium between gas condensation and evaporation.

To measure the photoluminescence spectra, we directly excited the designer nanodiamond with a green laser via free space, and collected the photoluminescence from the SiV center via free space. Figure S7 shows the photoluminescence spectra as a function of time. It is the same as Fig. 3b, but with an extended wavelength range. As we can see in Fig. S7, besides the SiV photoluminescence, we observed cavity-enhanced SiN autofluorescence at each cavity resonance. We used the cavity-enhanced SiN autofluorescence signal to monitor the cavity resonance at each time point. Specifically, we first traced out the cavity resonance for those cavity modes that are far detuned from the SiV center. Since the cavity free spectral range was a constant, we could then calculate the cavity resonance for those modes that are close to the SiV center. This is how we plotted the white dashed lines indicating the cavity resonances in Fig. 3b of the main text.

We noted that the autofluorescence of the SiN is generally detrimental to our measurements since it decreases the signal-to-noise ratio. This is why we were not able to perform a second order correlation measurement for photoluminescence of the SiV center. The autofluorescence of the SiN can be shifted to an irrelevant wavelength range by using nitrogen-rich SiN, as has been demonstrated recently.[1]

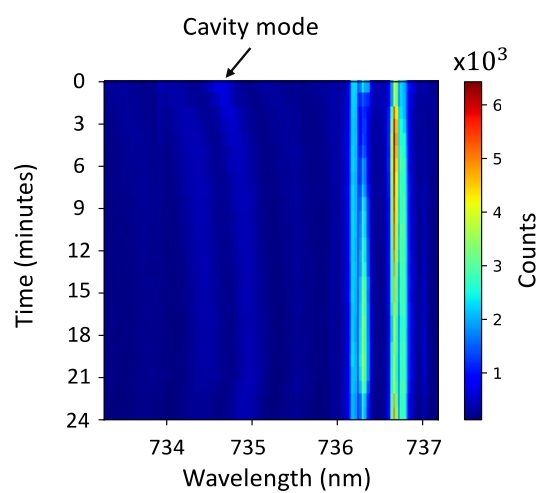

**Figure S7.** Figure 3b of the main text with extended wavelength range. The cavity-enhanced SiN autofluorescence (as indicated by the arrow) allows us to track the resonance of each cavity mode at every time point during the gas condensation.

## 6. Calculation of the lower bound of the Purcell factor

The Purcell factor for the SiV center coupled to the cavity can be calculated as[2]

$$F = \frac{\frac{\tau_{off}}{\tau_{on}} - 1}{\xi} + 1, \qquad (1)$$

where $\tau_{off} = 1.82 \pm 0.01$ ns is the lifetime of the SiV center when all the four transitions of the SiV center are far detuned from all cavity modes, $\tau_{on} = 1.70 \pm 0.01$ ns is the lifetime of the SiV center when the C transition is on resonance with one of the cavity modes, $\xi = \frac{\gamma_c}{\Gamma}$ is the ratio between the rate of spontaneous emission of the SiV center via the C transition, $\gamma_c$, and the total decay rate of the SiV center, $\Gamma$. To calculate the lower bound of the Purcell factor, we need to estimate the upper bound of $\xi$, which is simply a product of the quantum efficiency of the SiV center, the zero-phonon-line emission fraction, and the fraction of zero-phonon-line emission into the C transition. By assuming an upper bound of the SiV quantum efficiency of 30%, the zero-phonon-line fraction of 80%, and a fraction of 30% of zero-phonon-line emission into the C transition, we estimated the upper bound of $\xi$ to be $\xi = 0.072$, which leads to a lower bound value of the Purcell factor to be 2.

## 7. Calculation of the theoretically achievable Purcell factor

The theoretically achievable Purcell factor can be calculated as $F = \frac{3}{4\pi^2} \frac{Q}{V_{mode}} \left(\frac{\lambda}{n}\right)^3$, where $Q$ is the cavity Q factor, $V_{mode}$ is the cavity mode volume, calculated to be $V_{mode} = 28 \left(\frac{\lambda}{n}\right)^3$ based on FDTD, and $n = 2$ is the refractive index of silicon nitride. For the specific device, we reported the cavity Q factor of $Q_{loaded} = (1.01 \pm 0.06) \times 10^4$, measured when no gas was condensed on the cavity. However, the maximum lifetime reduction occurred when we tuned the cavity to be resonant with the C transition of the SiV center, at which the gas condensation employed for cavity tuning degraded the cavity Q to be ~3,000, measured from the cavity enhanced SiN autofluorescence as shown in Fig. S7. Based on these values, we calculated the theoretically achievable Purcell factor to be $F = 8$. This value is 4 times larger of the experimentally measured value. We attribute this mismatch to the misalignment of the polarization and position between the SiV center and the cavity mode.

## 8. Calculation of achievable photon extraction efficiency in our platform

The photon extraction efficiency, defined as the probability of extracting a spontaneously emitted photon into the cavity mode per SiV excitation, is about 6.5% for our current device. This number is currently limited by the constraints in our experimental measurements. Specifically, the gas condensation employed for cavity resonance tuning degraded the cavity Q factor. In the main text, we reported the cavity Q factor of $Q_{loaded} = (1.01 \pm 0.06) \times 10^4$, measured when no gas was condensed on the cavity. However, the maximum lifetime reduction occurred when we tuned the cavity to be resonant with the C transition of the SiV center via gas condensation, at which the Q factor was degraded to be ~3,000, measured from the cavity enhanced SiN autofluorescence as shown in Fig. S7. If we can tune the cavity and SiV center in resonance without degrading the cavity Q factor, we can achieve a photon extraction efficiency of 26% even with our current device.

The photon extraction efficiency can be further improved via better fabrication and the use of smaller designer nanodiamonds. The cavity Q factor achieved in our device, even for the unloaded cavity, is only $Q_{unloaded} = (1.17 \pm 0.09) \times 10^4$. This number is about two orders of magnitude lower than the Q factors achieved with the state-of-the-art SiN micro-ring resonators,[3] due to the limited access we have to high-quality etchers. If we can improve our SiN fabrication techniques, we should be able to easily reach the regime where the Q factor of the loaded cavity, $Q_{loaded}$, is capped by the integration of the designer nanodiamonds, which was calculated to be $Q_{hy} = (4.7 \pm 0.3) \times 10^4$ in the main text. Furthermore, we can further improve $Q_{hy}$ by using smaller designer nanodiamonds, and thus improve the photon extraction efficiency. The red line in Fig. S8 shows the calculated photon extraction efficiency as a function of the diameter of the integrated designer nanodiamonds. In this calculation, we assumed the designer nanodiamonds to have a cylindrical shape with a fixed height of 100 nm, which is reasonable to fabricate. We also assumed a uniform distribution of the possible SiV center position within the volume of the designer nanodiamonds, and a uniform distribution of the angle misalignment between the SiV center dipole moment and the cavity polarization from -45 degrees to 45 degrees. These assumptions ensured that the photon extraction efficiencies we calculated are achievable in the near-term experiments. As shown in Fig. S8, we can achieve a photon extraction efficiency of 97% with a designer nanodiamond of a diameter of 50 nm.

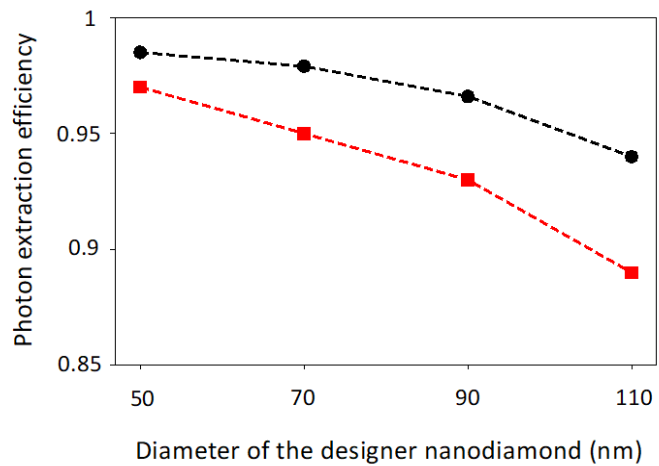

**Figure S8.** Calculated photon extraction efficiency as a function of the diameter of the integrated designer nanodiamond for both the micro-ring resonator (red line) and the fishbone photonic crystal cavity (black line).

Another way to improve the photon extraction efficiency is to employ cavities with much smaller mode volumes, such as a photonic crystal cavity. However, experimentally, it may be challenging to integrate the designer nanodiamonds with the air-hole-based photonic crystal cavity, since the distance between two neighboring air holes is typically comparable or even smaller than the size of the designer nanodiamonds we can make. For this reason, we have designed a fishbone photonic crystal cavity as shown in Fig. S9. This cavity possesses a calculated loaded Q factor of $Q_{loaded} = 6.4 \times 10^4$ and a mode volume of $3.7(\lambda/n)^3$, where $\lambda$ is the cavity resonance wavelength (in vacuum) and $n$ is the refractive index of SiN. The black line in Fig. S8 shows the calculated photon extraction efficiency as a function of the diameter of the designer nanodiamond based on the fishbone photonic crystal cavity. We made the same assumptions about the designer nanodiamonds height, the position mismatch, and the polarization mismatch as in the previous calculation. As shown in Fig. S8, we can achieve an efficiency of 98.5% with a designer nanodiamond of a diameter of 50 nm inside a fishbone photonic crystal cavity.

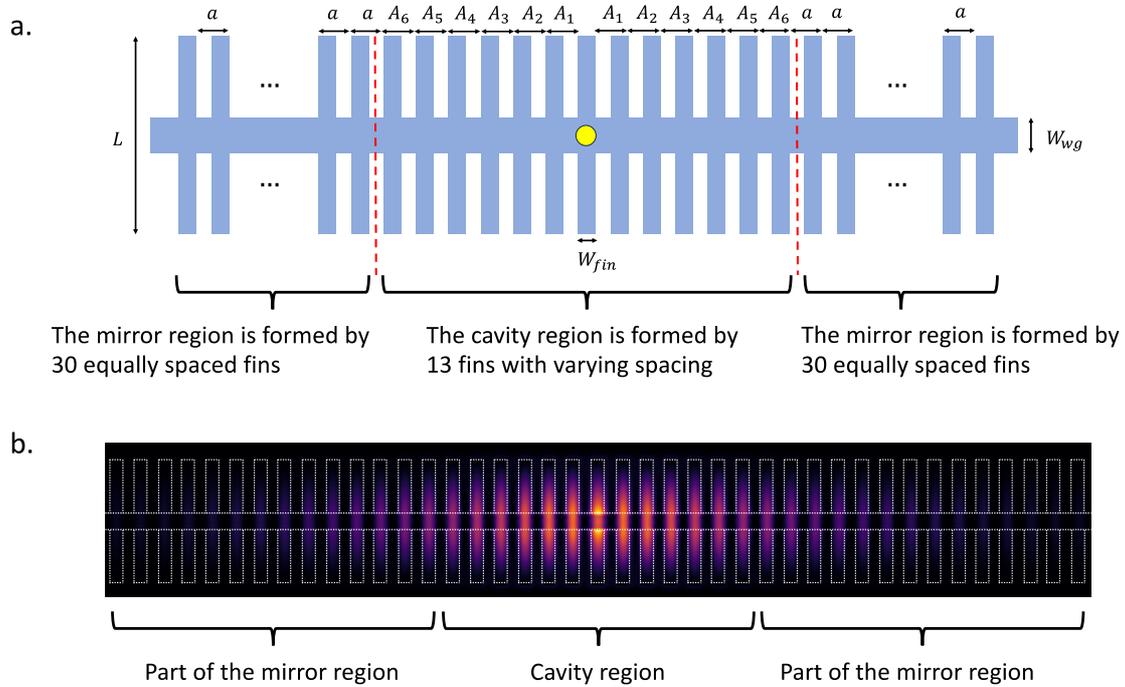

**Figure S9.** Design of the fishbone photonic crystal cavity. (a) The cavity is made of SiN with a thickness of 305 nm, sitting on top of a $SiO_2$ buffer layer. The photonic crystal is formed by fins with a length of $L = 1217\ nm$ and a width of $W_{fin} = 130\ nm$. The waveguide width is $W_{wg} = 110\ nm$. The mirror region of the photonic crystal is formed by 30 fins on each side with a lattice constant of $a = 213\ nm$. The cavity region is formed by 13 fins in total, with a modulated lattice constant of $A_k = a\left(1 - 0.03\left|2\left(\frac{k}{7}\right)^3 - 3\left(\frac{k}{7}\right)^2 + 1\right|\right)$,[4] where $A_k$ is the lattice constant of the $k$-th fin counted from the center of the cavity, and $k$ goes from 1 to 6. In the calculation of the photon extraction efficiency, we assumed that a designer nanodiamond was placed at the center of the cavity, as indicated by the yellow circle. (b) Mode profile of the cavity calculated by FDTD.